\documentclass[8.5pt,twoside,twocolumn]{article}

\usepackage[super,sort&compress,comma]{natbib} 
\usepackage{mhchem}
\usepackage{times,mathptmx}
\usepackage{sectsty}
\usepackage{balance} 
\usepackage{natbib}
\usepackage{hyperref}
\hypersetup{colorlinks=true,linkcolor=blue,citecolor=blue,urlcolor=blue}
\usepackage{graphicx} %eps figures can be used instead
\usepackage{lastpage}
\usepackage[format=plain,justification=raggedright,singlelinecheck=false,font=small,labelfont=bf,labelsep=space]{caption} 
\usepackage{fancyhdr}
\usepackage{setspace}
\begin{document}

\thispagestyle{plain}
\fancypagestyle{plain}{
\renewcommand{\headrulewidth}{1pt}}
\renewcommand{\thefootnote}{\fnsymbol{footnote}}
\renewcommand\footnoterule{\vspace*{1pt}% 
\hrule width 3.4in height 0.4pt \vspace*{5pt}} 
\setcounter{secnumdepth}{5}

\makeatletter 
\def\subsubsection{\@startsection{subsubsection}{3}{10pt}{-1.25ex plus -1ex minus -.1ex}{0ex plus 0ex}{\normalsize\bf}} 
\def\paragraph{\@startsection{paragraph}{4}{10pt}{-1.25ex plus -1ex minus -.1ex}{0ex plus 0ex}{\normalsize\textit}} 
\renewcommand\@biblabel[1]{#1}            
\renewcommand\@makefntext[1]% 
{\noindent\makebox[0pt][r]{\@thefnmark\,}#1}
\makeatother 
\renewcommand{\figurename}{\small{Fig.}~}
\sectionfont{\large}
\subsectionfont{\normalsize} 

\fancyfoot{}
\fancyfoot[RO]{\footnotesize{\sffamily{1--\pageref{LastPage} ~\textbar  \hspace{2pt}\thepage}}}
\fancyfoot[LE]{\footnotesize{\sffamily{\thepage~\textbar\hspace{3.45cm} 1--\pageref{LastPage}}}}
\fancyhead{}
\renewcommand{\headrulewidth}{1pt} 
\renewcommand{\footrulewidth}{1pt}
\setlength{\arrayrulewidth}{1pt}
\setlength{\columnsep}{6.5mm}
\setlength\bibsep{1pt}

% Author/Title section .........

\twocolumn[
\begin{@twocolumnfalse}
\noindent\LARGE{\textbf{Proximity induced Colossal Conductivity Modulation in Phosphorene}}
\vspace{7pt}

\noindent\large{\textbf{A. Chaudhury, S. Majumder, S. J. Ray$^{\ast,1}$}}

\vspace{0.6cm}

%\date{\today}

\noindent \normalsize{Phosphorene is a promising single elemental two-dimensional layered semiconductor with huge potential for future nanoelectronics and spintronics applications. In this work, we investigated the effect of an organic molecule (benzene) in the close proximity of a Phosphorene nanoribbon. Our extensive calculations reveal that the semiconducting nature of Phosphorene stays unaffected as a result of the molecular adsorption while the transport properties go through drastic changes. Under the influence of dopant atoms and external strain, colossal changes in the conductivity is observed with a maximum enhancement $>$ 1500\% which has not been observed earlier. This effect is pretty robust against the (i) variation of system size, (ii) type, location and concentration of dopants and (iii) nature and magnitude of the external strain. Furthermore, we demonstrated how a gate voltage can be used to fine tune the enhanced conductivity response in a Field-effect transistor (FET) structure. Our results provide new direction for Phosphorene based nanoelectronics in applications like sensing, switching where a higher level of conduction can offer better resolution, higher ON/OFF ratio and superior energy efficiency.}
\vspace{0.5cm}
\end{@twocolumnfalse}
]

\footnotetext{\textit{Department of Physics, Indian Institute of Technology Patna, Bihta 801106, India; E-mail$^1$: ray@iitp.ac.in, ray.sjr@gmail.com}}

%\keywords{ Phosphorene, zigzag and armchair direction, anisotropy, sensor, strain, doping}

%\maketitle
%\setcounter{figure}{0}

\section{Introduction}

Two dimensional (2D) layered structures are crystalline nanomaterials with superior electronic, optical, mechanical properties that have potential applications in future nanoelectronics and spintronics \cite{NovoselovPNAS}. Doping in such materials is commonly used for tuning the band gap and electronic properties, designing of high performance sensor, pn-junction, enhancement of the conductivity in a FET structure etc. while presence of various transition metal atoms can offer ferromagnetic functionalities and diverse magnetic ground states which are useful in designing various spintronic devices. Among the recently discovered 2D materials, Phosphorene is anticipated to be a preferred candidate over graphene and transition metal dichalcogenides for future nanoelectronics as it uniquely offers ambipolar transport \cite{Das2014}, high carrier mobility \cite{Long2016}, large ON/OFF ratio \cite{Li2014, Liu2014}, tuneable band gap and low spin-orbit coupling \cite{Avsar2017, Kamalakar2015}. Additionally application of strain on Phosphorene results in conduction anisotropy \cite{Fei2014, Rodin2014, Ray2018}, phase transition \cite{Rodin2014, Guan2014, Wu2015, Elahi2015, Peng2014, Han2014}, tuneable Young's modulus \cite{Wei2014}, negative Poisson's ratio \cite{Jiang2014}, bandgap engineering \cite{Guan2014, Han2014, Zhu2014, Deniz2014, Guo2014, Li2014c, Qiao2014, Rudenko2014, Tran2014, Dai2014, Cai2014}, Dirac-like cones \cite{Can2015, fei2015} etc. which offers a new way to manipulate the structure as per the requirement. Owing to these excellent attributes, intense research in Phosphorene is in full swing to look for new phenomena and applications that are continuing to fascinate scientists and technologists.

Due to the enhanced surface to volume ratio, 2D materials are sensitive to the presence of a neighbouring molecule as demonstrated in graphene\cite{Schedin2007} and carbon-nitrogen compounds \cite{Rani2018}. The adsorption process results in charge transfer between the molecule and the host layer leading to an enhanced conduction. From a nanoelectronic perspective, enhanced conduction is very much desirable in designing a Field effect transistor (FET) or similar switching elements\cite{Ray2014a, Ray2014b, Ray2014c}, where the increase in the ON current can significantly improve the ON/OFF ratio and the switching performance. Due to constant size scaling in the CMOS-architecture, presence of leakage current is a major issue of concern. Enhancement of the conduction in 2D materials through a reversible route can significantly improve the signal to noise ratio and better contrast in performance at a higher energy efficiency.

In the present work, we have used First-principles calculations to study the effect of the proximity of an organic molecule - Benzene (C$_6$H$_6$) on the transport and electronic properties of an armchair Phosphorene nanoribbon (APNR). Benzene lies at the heart of organic chemistry and is widely used in various industries due to its easy availability. Using Nonequilibrium Green's Function (NEGF) formalism, the current-voltage characteristics of the APNR was studied in 2-probe geometry in the presence and absence of the Benzene molecule. Furthermore, we investigated the effect of the system size, doping of varying types (p/n-type) and location on the conductivity response and how it can be tuned using uniaxial strains. Under specific conditions of strain and doping, an enormous increase ($> 1500\%$) in the current signal was observed indicating the superior enhancement of conductivity response of Phosphorene in the presence of such molecule. Furthermore, in a realistic FET geometry, we have explored how a gate voltage can be used to fine-tune the conductivity behaviour. We believe that our results will provide detailed insight in the tuning and control  of the conductivity of Phosphorene in a sub-10 nm structure, finding desirable conditions to unveil the most prudent response that will lead to new possibilities in 2D nanoelectronics.
\begin{figure}
\includegraphics[width=7cm]{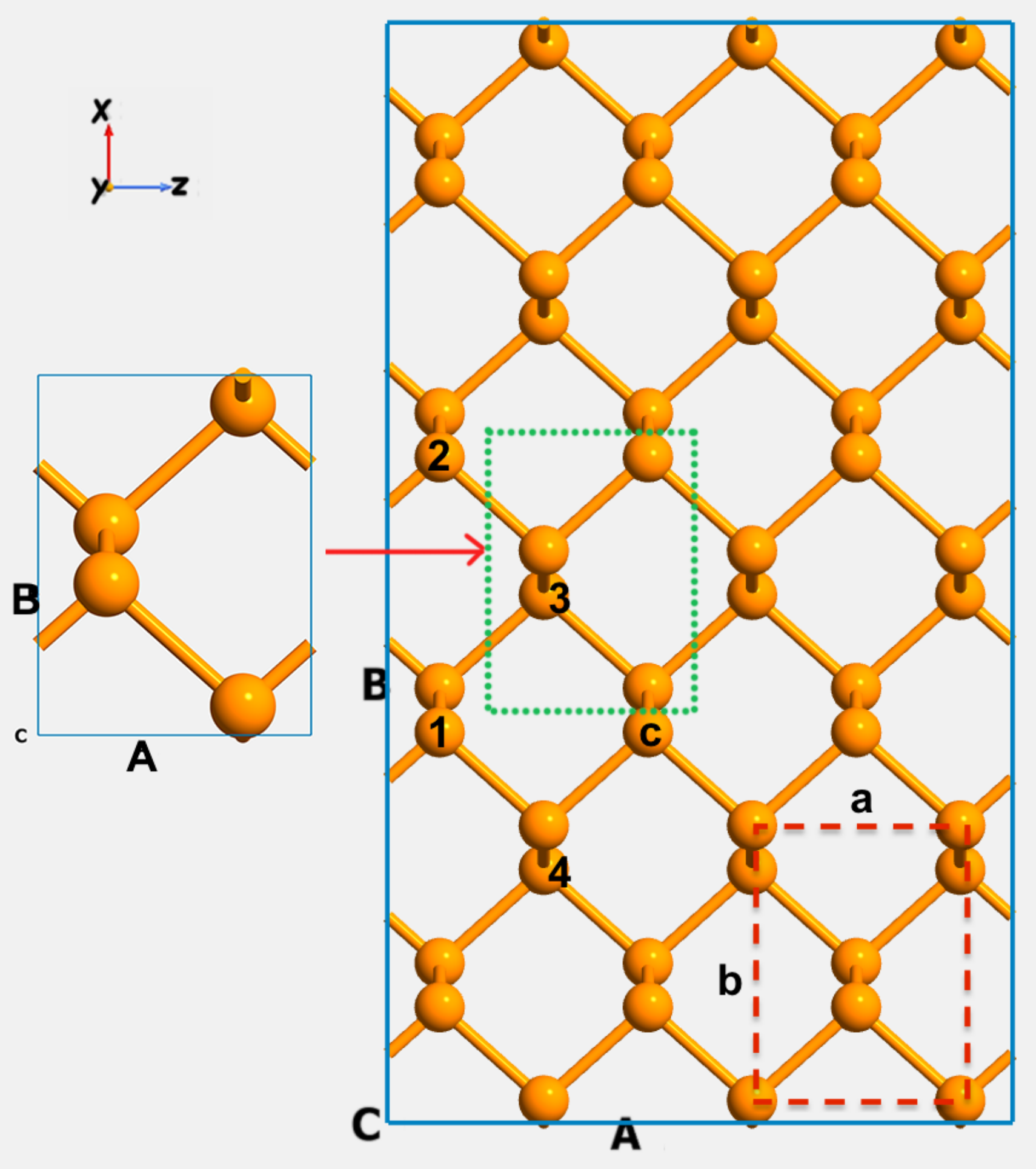}
\caption{{\small Schematic of the top view of (left) unit cell of 2D phosphorene and (right) 3$\times$4 supercell of 2D phosphorene where the two dotted boxes are the unit cells with \textbf{a} and \textbf{b} as the lattice vectors. Various doping locations are marked by C (centre doping) and 1, 2, 3, 4 for the locations for edge doping with with those numbers of dopant atoms.}}
\label{fig.1}
\end{figure}

\section{Computational details}

First-principles calculation of the electronic structure and geometry relaxation have been carried out by self-consistent density functional theory using the Atomistix ToolKit (ATK)\cite{ATK}. The exchange-correlation energy was assessed using the Perdew-Burke-Ernzerhof  exchange-correlation functional \cite{PBE} within the Generalised Gradient Approximation (GGA), where the wave functions are expanded within double-$\zeta$ polarized basis set under the Periodic boundary conditions with an energy cut-off limit of 180 Ry. The reciprocal space of the Brillouin Zone was sampled using a 27 $\times$ 27 $\times$ 1 Monkhorst-Pack $k$-grid\cite{Monkhorst}. To reduce the interactions between neighbouring layers, a minimum vacuum space of 15$\rm{\AA}$ was used in the non-periodic directions.  The APNR structures (in pristine and adsorbed conditions) were structurally relaxed until the force on each atom was less than 10$^{-3}$ eV/\AA \,in the equilibrium condition.

All the transport calculations in the 2-probe configuration were carried out using the NEGF  combined Density Functional Theory methodology\cite{Soler2002, Brandbyge2002}. The current was estimated using the Landauer-B$\ddot{u}$ttiker formula \cite{Landauer, Butikker} given by,
\begin{equation}
I = \frac{2e}{\hbar}\int_{-eV/2}^{eV/2}T(E,V)[f_L(E-\mu_L) - f_R(E-\mu_R)]dE
\label{eqn.1}
\end{equation}
where T(E,V) is the transmission function at an energy E, bias voltage V and $f_{L/R}(E,\mu_{L/R})$ is the Fermi distribution and $\mu_\mathrm{L}$ and $\mu_\mathrm{R}$ are the electrochemical potentials of the left and right electrodes respectively. Details of the methodology used for the calculation of T(E,V) and $\mu_\mathrm{L}$ and $\mu_\mathrm{R}$ are described in the supplementary information section\cite{SI-Phosphorene-transport}. A $k$-point sampling of (1 $\times$ 1 $\times$ 300) has been adopted for the transport calculations (300 along the transport direction) as no changes in the transmission were observed beyond this limit.

\begin{figure*}
\includegraphics[scale=.6]{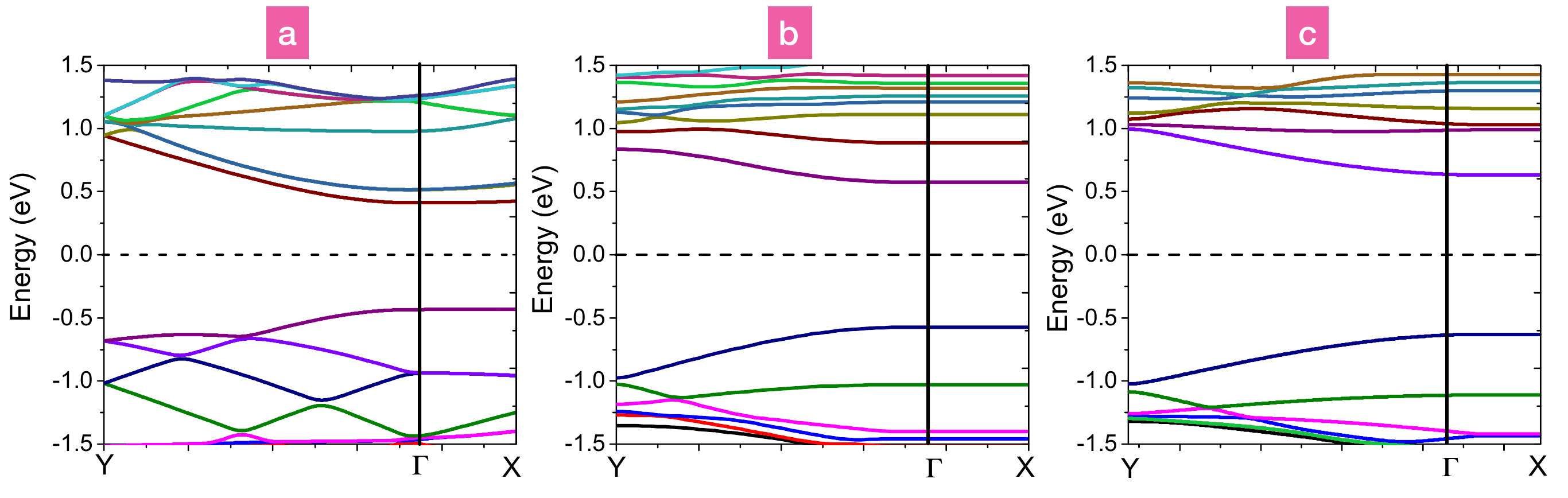}
\caption{{\small Band structure of $\mathrm{A_{34}}$ in (a) unpassivated, (b) H-edge passivated and (c) after absorbing benzene molecule configurations.}}
\label{fig.2}
\end{figure*}

\section{System Description}

The unit cell of Phosphorene (Fig.~\ref{fig.1}) has lattice constants \textbf{a} = 3.3136 \AA, \textbf{b} =  4.3763 \AA, where \textbf{a} and \textbf{b} are the distances between two consecutive repeating P atoms along the zigzag and armchair directions respectively \cite{Ray2016a, Ray2016b}. The bond length along the zigzag direction containing the in-plane hinge angle of 95.370$^{\circ}$ is 2.24 \AA\, while the connecting bond lengths along the armchair direction is 2.26 \AA. We designated an $m\times n$ sized APNR as $\mathrm{A_{mn}}$ which has dimensions : $m$ times primitive lattice parameter, \textbf{a} along the zigzag direction and $n$ times the primitive lattice parameter \textbf{b} along the armchair direction. In the present work, we have considered a H-edge passivated $3\times4$ supercell of APNR as our central structure, which is abbreviated as $\mathrm{A_{34P}}$ in pristine structure (no doping and defect) and $\mathrm{A_{34A}}$ as the (benzene) adsorbed configuration while $\mathrm{A_{34}}$ is used for general use.

\section{Results and Discussion}

\subsection{Pristine System}

The electronic band structure of $\mathrm{A_{34}}$ in unpasivated, passivated and benzene adsorbed condition is illustrated in Fig.~\ref{fig.2}, which is observed to be a direct band gap semiconductor occourring at the $\Gamma$ point. Compared to the unpassivated system, an increment in bandgap is observed from 0.88 eV to 1.24 eV due to H-passivation. The origin behind this increment can be traced by comparing the band structures between these two cases. As shown in Fig.~\ref{fig.2}(b), the valence band shifted downwards whereas the conduction band moved upwards in energy after passivation, thus increasing the bandgap of $\mathrm{A_{34}}$. After absorbing Benzene molecule the bandgap of $\mathrm{A_{34}}$ changes to 1.26 eV (Fig.~\ref{fig.2}c). This suggests that the semiconducting nature of the APNR stays unaffected through the process of molecular adsorption. Experimentally reported band gap for monolayered phosphorene lies between 0.98 - 1.45 eV \cite{Das2014, Liu2014, Wang2015b}, which is comparable to our estimated value. With a change in the band gap, the barrier heights for electrons and holes will vary depending on the relative positions of the conduction and valance bands with respect to the electrodes. Transmission will occour at those energies outside the band gap, depending on the applied bias and relative positions of Fermi levels of left and right electrodes.
\begin{figure}
\includegraphics[width=8cm]{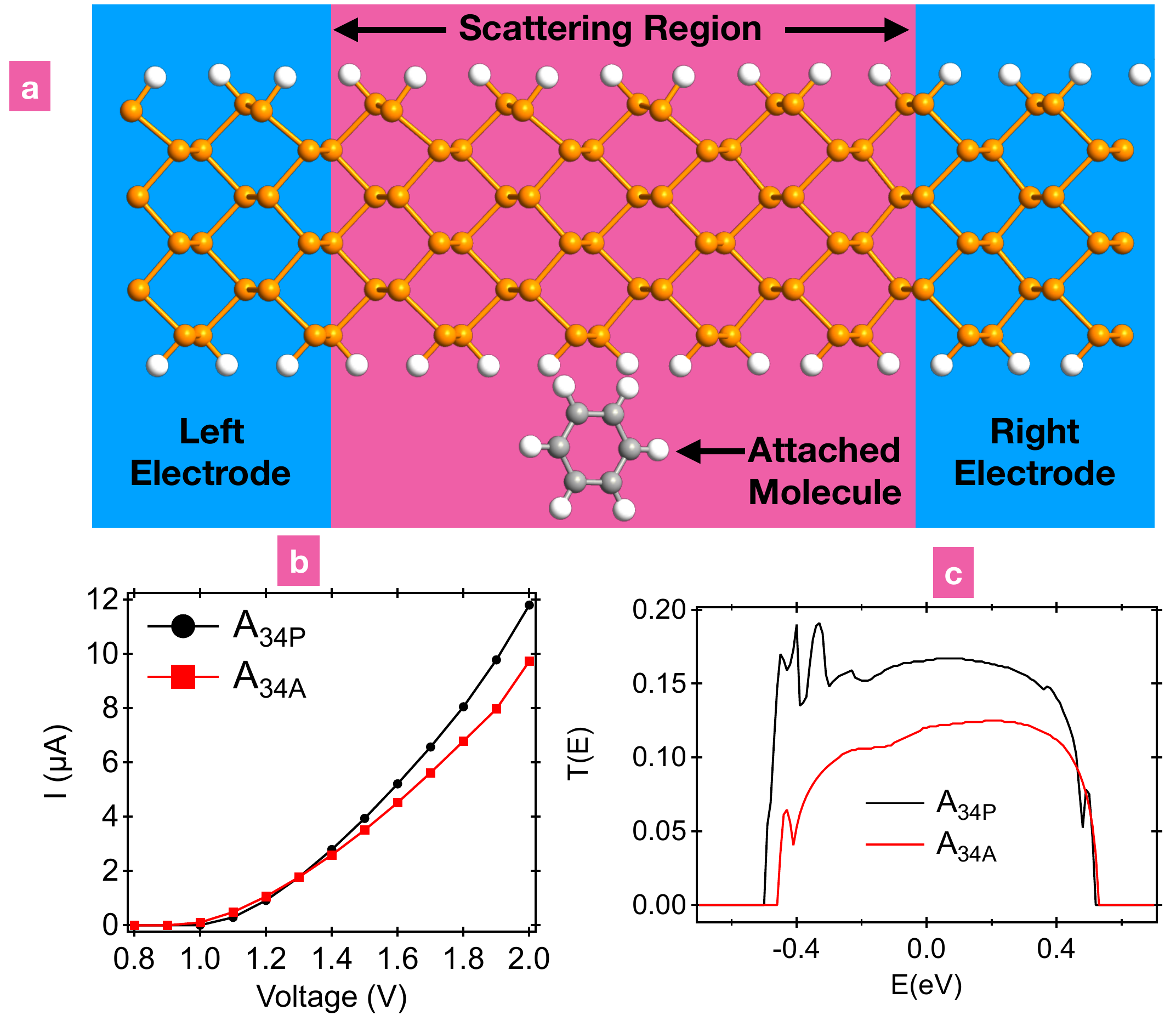}
\caption{{\small (a) Schematic of an APNR in 2-probe geometry for IV-calculations with attached Benzene molecule, (b) Current-voltage relationship of $\mathrm{A_{34}}$ in 
pristine and adsorbed configuration, (c) Corresponding transmission spectrum at 2.0V of applied bias.}}
\label{fig.3}
\end{figure}

In order to measure the current-voltage characteristics of $\mathrm{A_{34}}$, the central scattering region is connected between two semi-infinite perfect electrodes as shown in Fig.~\ref{fig.3}(a). In this measurement geometry, benzene molecule was adsorbed to the side of the APNR. Out of the several configurations tested in this geometry, it was observed that the planar configuration (molecule and APNR in the same plane) has a lower energy compared to the tilted configuration (molecule staying at angle with respect to the APNR) as shown in Fig.~S1 (Supporting Information/SI)\cite{SI-Phosphorene-transport}. Apart from the side adsorption scenario, we have also considered the situation with the molecule adsorbed on top of the layer at various locations. However, the planar configuration as illustrated in Fig.~\ref{fig.3}(a) is found to be the optimal adsorption configuration and thus considered as the preferred geometry here.

\begin{figure*}
\includegraphics[width=18cm]{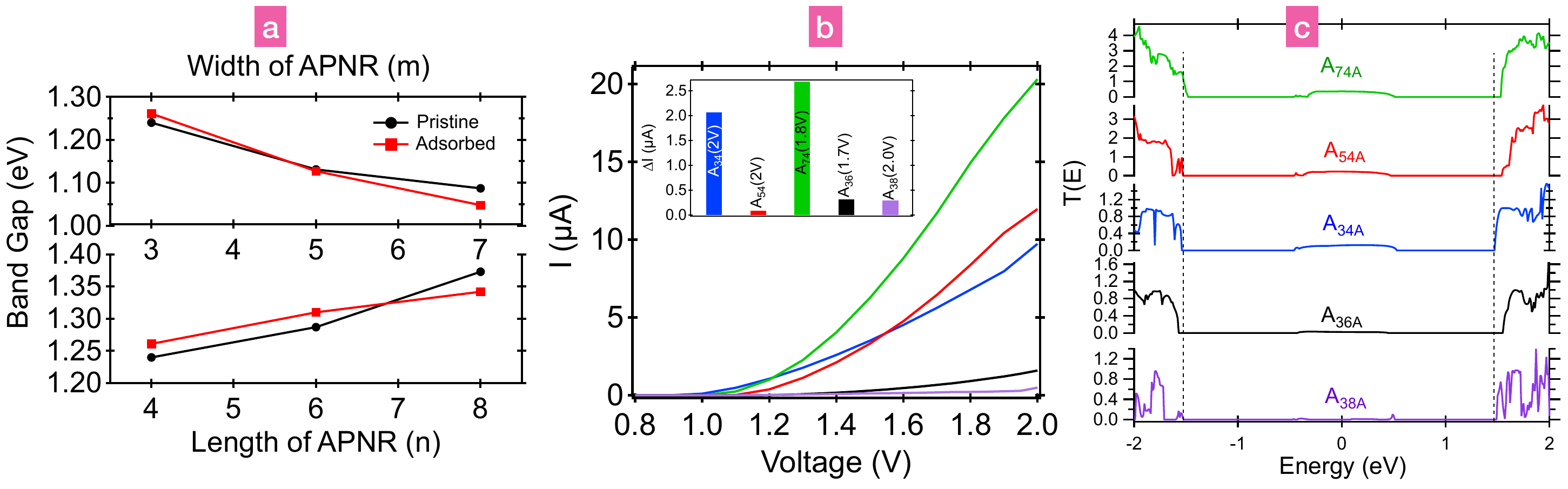}
\caption{{\small (a) Dependance of the Bandgap with the length ($n$) and width ($m$) of APNR in pristine and adsorbed configuration, (b) Current-voltage relationship of $\mathrm{A_{mn}}$ in the adsorbed configuration and (inset) maximum $\Delta$I for these cases, (c) Transmission spectrum at 2.0V applied bias for these adsorbed configurations respectively.}}
\label{fig.4}
\end{figure*} 

The IV-characteristics of $\mathrm{A_{34}}$ is shown in Fig.~\ref{fig.3}(b). At zero-bias no current flows as the density of states in both the electrodes are identical. A non-zero value of current is observed when the applied bias goes above 0.9 V, which roughly corresponds to the band gap of $\mathrm{A_{34}}$. In the presence of the attached molecule, the overall shape of the IV-curve shows the characteristics of a gapped structure as also observed in the pristine case. This agrees well with the trend observed from the band structure calculations. However, it is noticed that attaching the molecule results in a small reduction of current at around 1.2V and the difference increases towards higher applied bias. This can be explained as follows : the conduction pathways are lying along the central part and edges of the layer. When the molecule is adsorbed, the local density of electronic states along the conduction paths decreases. Thus the path available for conduction and the transmission gets slightly reduced, resulting in a reduced current/conduction. This can be confirmed from the transmission behaviour as illustrated in Fig.~\ref{fig.3}(c) where the value of transmission between the energy range of -0.5 eV to 0.5 eV is clearly higher for the pristine configuration compared to the adsorbed counterpart. As the IV-characteristics in the two cases is significantly different at higher voltages, this suggests that APNR is sensitive to the presence such organic objects and the sensitivity can be measured from the difference of current(s) at a particular applied bias. A quantitative estimation can be done using the modulation factor ($\eta$) defined at a given bias V as,
\begin{equation}
\rm{\eta(V) = \Bigg|\frac{I_{A}(V) - I_{U}(V)}{I_{U}(V)}} \Bigg|= \Bigg|\frac{\Delta I (V)} {I_{U}(V)}\Bigg|
\end{equation}
where $I_A$ and $I_U$ are the currents in adsorbed and unadsorbed configuration of the APNR respectively.

Next, we have checked the effect of system size on the electronic and transport properties in the presence of the molecule. The size dependence of the band gap as function of length ($n$) and width ($m$) is illustrated in Fig.~\ref{fig.4}(a). With an increase in width, the band gap gets reduced which is a consequence of quantum confinement effect. The band gap on the other hand increases with an increase in the length ($n$). Also, the bands become flatter (Fig.~S2 in SI\cite{SI-Phosphorene-transport}) as the length increases. It is a result of increased scattering along the ribbon length causing the transmission probability to decrease drastically, which implies an increase in effective mass and hence the flatter bands. However, the band dispersions remain more or less same with $m$ variation. In both the length and width variation, the band gap for the adsorbed system follows the similar trend as that of its pristine counterpart. 

The IV-relationships in the adsorbed configuration for various combinations of $m, n$ are illustrated in Fig.~\ref{fig.4}(b). With an increase in $m$, the current at higher bias gets enhanced while the reverse trend is observed with an increase in $n$ i.e the reduction in current. The highest current is observed when the band gap is the lowest. With an increase in the length of the ribbon, the overall value of transmission decreases as also shown in Fig.~\ref{fig.4}(c). For A$\rm{_{34A}}$,  A$\rm{_{54A}}$ and A$\rm{_{74A}}$ a non-zero transmission peak around 0 eV is visible which almost disappears for A$\rm{_{36A}}$ and A$\rm{_{38A}}$. With an increase in $m$, the central window of transmission stays more or less unchanged as represented through the dotted lines. But the actual value of transmission increases manifold at other energies. As the calculated current is the integral sum of transmission, hence this gets reflected through the IV-patterns (current density in Fig. S11 in SI\cite{SI-Phosphorene-transport}). While the width is kept fixed, but length is increased - the carriers have to travel a larger path and hence probability of scattering increases and transmission get reduced. On the contrary, for a fixed length an increase in the ribbon width can add additional transmission pathways that can result in an enhanced conduction. While looking at the change in current ($\Delta$I) as a result of the adsorption, a measurable difference is observed for all the configurations tested here as displayed in Fig.~\ref{fig.4}b (inset). A maximum in $\Delta$I $\sim$ 2.7 $\mu$A is observed for A$\rm{_{74P}}$ which is about 29.5\% higher than the value estimated for A$\rm{_{34P}}$. Such values of $\Delta$I are experimentally measurable and well-above the experimental noise floor ($\sim$ pA) at room temperature and much higher than the typical values reported for Phosphorene \cite{Kou2014b, Xu2015, Maity2016}. In terms of modulation, $\eta \sim$ 150\% is observed for A$\rm{_{38P}}$, which indicates the large change of the transport properties of APNR can be brought in the presence of the molecule. This property can thus be well-utilised for designing a nano-sensor made of Phosphorene. With a change of the edge structure, it will be interesting to study the behaviour of zigzag nanoribbon under identical ciscumstances. In earlier works involving graphene nanoribbon, it was observed that transport behaviour gets modulated independent of the edge structures \cite{Chowdhury2011}. The shape of the nanoribbon determines the transport pathways for conduction and any modification to this can perturb the other. Passivation of the edge structure offers better stability compared to an unpassivated system due to the saturation of the dangling bonds.

\subsection{Effect of Doping}
\begin{figure}
\includegraphics[scale=.35]{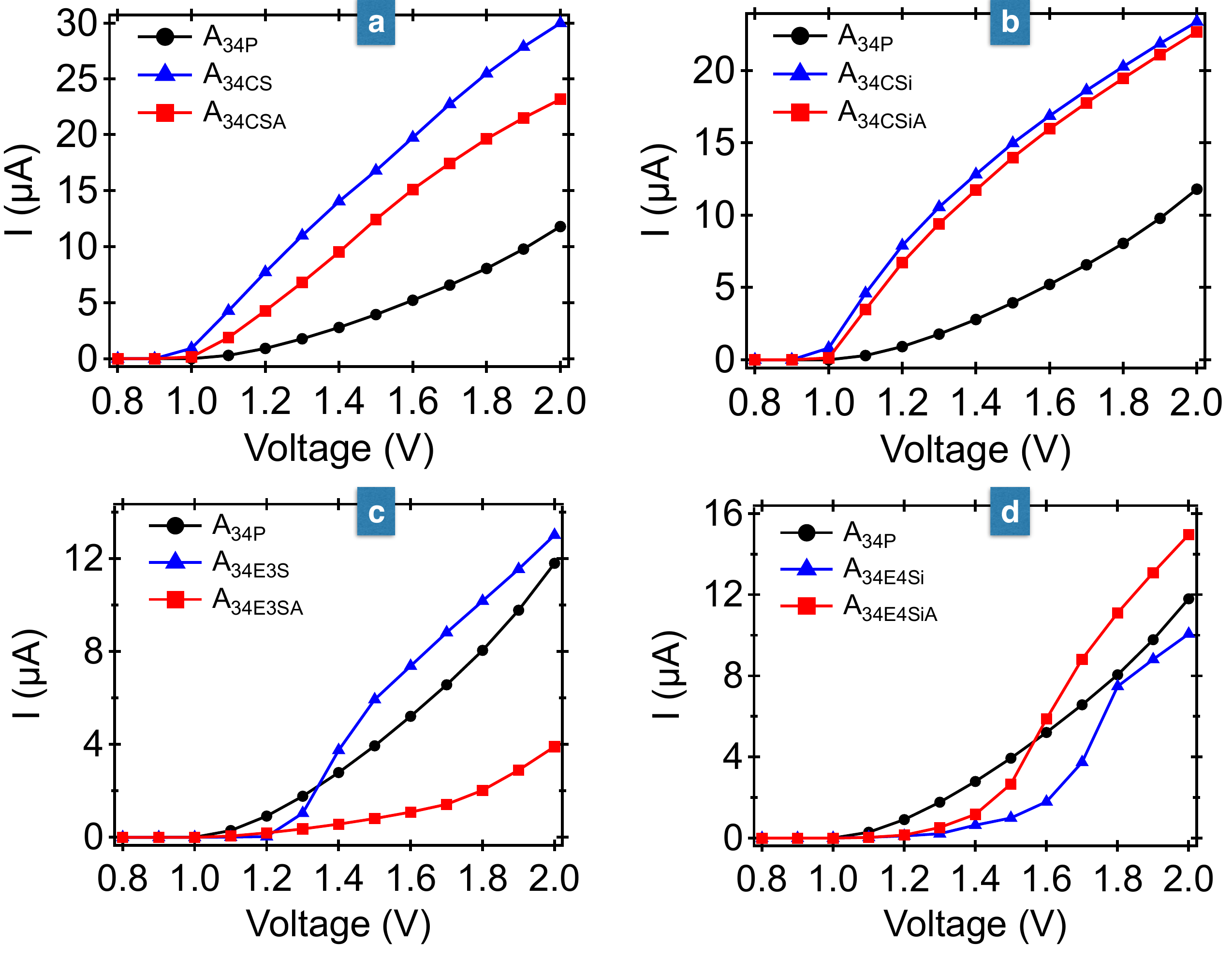}
\caption{{\small IV-characteristics of $\mathrm{A_{34}}$ in doped configurations with (a) centre-doped with S, (b) centre-doped with Si, (c) edge-doped with 3 S atoms,  and (d) edge-doped with 4 Si atoms. The red, blue and black lines represent the undoped, doped and adsorbed configurations respectively.}}
\label{fig.5}
\end{figure}
To check the effect of doping on the electronic and transport properties of the APNR, two different doping strategies were employed. Silicon was used as p-type dopant while Sulphur as n-type dopant. For both types, site specific doping was performed by placing the dopant atom at the centre (centre doping) and at the edge (edge doping). In our calculation edge doping concentrations have been increased gradually by sequentially replacing the numerically marked sites (see Fig.~\ref{fig.1}) with 1(1.56\%); 2(3.125\%); 3(4.59\%) and 4(6.25\%) dopant atoms. They are named as C (centre doped with 1 atom) and E1-E4 (edge doped with 1-4 atoms respectively). The nomenclature used to define these doping configurations is the following : $\mathrm{A_{34CSi}}$ represents centre doping with Si of $\mathrm{A_{34P}}$ while $\mathrm{A_{34CSiA}}$ denotes the same in the adsorbed configuration. Similarly, $\mathrm{A_{34E3S}}$ is used for Edge doped $\mathrm{A_{34P}}$ with 3 S atoms and $\mathrm{A_{34E3SA}}$ for the corresponding adsorbed structure. For $\mathrm{A_{34P}}$, edge doping is limited to a maximum of 4 dopant atoms to prevent doping of the electrodes, which is undesirable. With an increase in the n-type doping concentration, the Fermi level shifts downward whereas for p-type doping Fermi level shifts towards the higher energies (see Fig.~S3 in SI\cite{SI-Phosphorene-transport}). 

The IV-curves for $\mathrm{A_{34P}}$ centre doped with Sulphur is illustrated in Fig.~\ref{fig.5}(a). In all the 3 configurations, significant current starts flowing when the applied bias crosses 0.9V, which is a signature of a semiconductor with a sizeable bandgap. Compared to the pristine configuration, the value of current stays significantly higher in the doped geometry and the difference gets enhanced ($\rm{\Delta I_{max} = 18.16 \mu A}$) with an increase in the bias. In the proximity of the benzene molecule, the IV-pattern follows a similar trend as that of the doped case but a marked difference between the two lines is significant for the entire bias region. In this doped configuration, a maximum value of $\eta \sim$ 83.57\% is observed and a maximum $\rm{\Delta I = 6.8\mu A}$. In a similar manner pronounced effect of Si-doping at the centre can be seen from the IV-characteristics in Fig.~\ref{fig.5}(b). With an increase in the applied bias above the gap size of APNR, current in the doped configuration gets highly enhanced compared to its pristine counterpart with a maximum $\rm{\Delta I = 10.9\mu A}$ at 2V. In the presence of benzene, the difference in the current between the two configurations is clearly visible.

\begin{figure*}
\includegraphics[scale=.6]{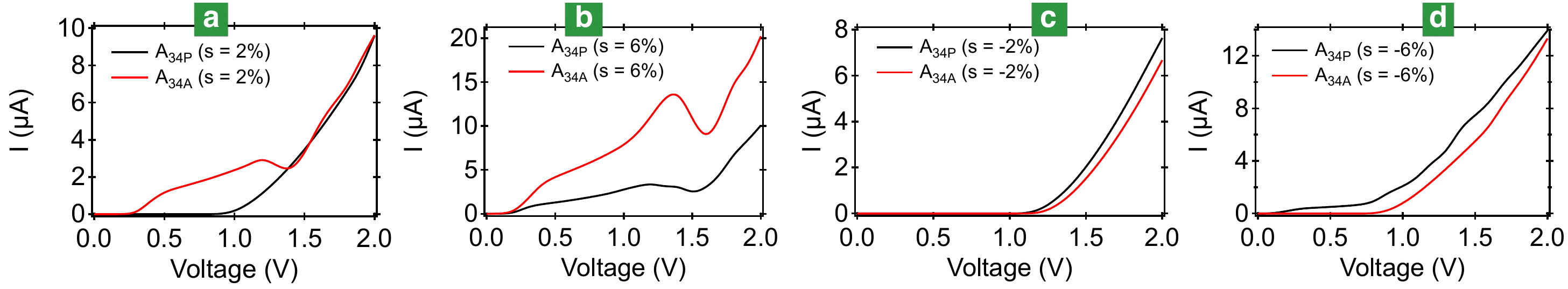}
\caption{{\small IV-response of $\mathrm{A_{34}}$ in pristine and adsorbed configurations under (a) 2\%, (b) 6\%, (c) -2\% and (d) -6\% applied strain.}}
\label{fig.6}
\end{figure*}

For the edge-doped configuration with 3 S and 4 Si atoms, the IV-characteristics are shown in Fig.~\ref{fig.5}(c-d) respectively. For both types of dopants, the onset voltage ($V_{onset}$) at which finite current starts flowing in the system gets enhanced in the doped configuration compared its pristine counterparts. For Si-doping in Fig.~\ref{fig.5}(d), $V_{onset} \sim$ 0.9V in the pristine structure while in the doped geometry, $V_{onset}\sim$ 1.2V. This trend is present for S-doping in Fig.~\ref{fig.5}(c), but the difference between the onset voltage between the pristine and doped configuration increases with an increase in the doping percentage. With an increment in the applied bias in Fig.~\ref{fig.5}(d), current starts rising monotonically in the doped configuration both in the presence and absence of the benzene molecule and clear difference between the current(s) indicates that edge-doping has a more pronounced effect for Si (Fig.~S4 in SI\cite{SI-Phosphorene-transport}). This leads to a huge enhancement $\eta$ of 228.07\% and a significant $\rm{\Delta I\sim 5\mu A}$ at high voltages. 

For S-doping at the edge in Fig.~\ref{fig.5}(c), a pronounced difference in the current was observed as a result of the proximity effect with a $\eta$ = 569.80\% and a maximum $\rm{\Delta I = 9.1\mu A}$. The large change in the current suggests the enhanced proximity influence of the C$_6$H$_6$ molecule in this APNR. In the edge doped APNR, additional carriers are locally more available near the edge region which is in closer proximity to the benzene molecule. This leads to a larger change in the transmission as shown in Fig.~S5 in SI\cite{SI-Phosphorene-transport}. The disappearance of the transmission peak around the Fermi level in the presence of Benzene significantly reduces the integrated sum of T(E) over the entire bias range leading to a huge reduction of current.  For various doping configurations, the values of $\eta$ are tabulated in Table~\ref{tab.1} which are significantly large independent of the doping type and configuration. The general trend suggests that both p/n-type dopants are useful for the conductivity modulation of APNR. As doping in a semiconductor introduces additional charge carriers for conduction, which for a 2D material with higher surface sensitivity results in an enhanced charge transfer between the layer and the molecule and conductivity changes. To better the value of $\eta$, edge doping is more effective which for Si increases monotonically with an increase in the doping percentage. This offers additional tuneability in the conduction response in the presence of an organic molecule.

\begin{table}
\caption{A comparison of the maximum conductivity modulation ($\eta$) of the APNR in various doping configurations}
\begin{spacing}{1.5}
\begin{center}
\begin{tabular}{c|c|c}\hline
Doping & $\eta$(Si) & $\eta$ (S)\\ 
Configuration & & \\\hline \hline
C1 (1.56\%) & 84.24 \%(1 V ) &  83.57\%
(1 V)\\
E1 (1.56\%) & 39.29\% (1.1 V) & 90.77\%
(1.1 V)\\
E2 (3.13\%) & 90.32\% (1.1 V) & 142.07\%
(1.1 V) \\
E3 (4.59\%) & 206.44\% (1.1 V) & 569.80\%
(1.2 V) \\
E4 (6.25\%) & 228.07\% (1.6 V) &  84.94\%
(1.1 V) \\
\hline
\end{tabular}
\end{center}
\end{spacing}
\label{tab.1}
\end{table}

The presence of Benzene molecule has not been found to affect the band structure of the doped APNR (both for Si or S doping). However, the transport properties gets affected to a great degree and showed varying characteristics depending upon the site and concentration of doping as the dopant states are induced near the Fermi level of the central scattering region. It is expected that in the case of $\mathrm{A_{34P}}$ where the current decreased on adsorption, similar circumstance would occur for the doped APNR until sufficient delocalised energy states are provided by the dopant atoms and in our study we also observed similar situation. We have noticed decrement in current after absorption of Benzene molecule when the APNR is doped upto 2 Si but when the system is doped with 3 and 4 Si atoms respectively current increased after the absorption of Benzene molecule. On the other hand for Sulphur doped APNR, fall in current is observed when the system is doped with 1 S atom (both centre and edge doped) and rise in current is observed after the system is doped with 2 or 4 S atom which satisfy our claim.

\begin{table}
\caption{A comparison of the maximum conductivity modulation ($\eta$) of the APNR at various values of strain percentages}
\begin{spacing}{1.8}
\begin{center}
\begin{tabular}{c|c||c|c}\hline
Strain & $\eta$ & Strain & $\eta$\\ \hline \hline
2\% & 1527.81 (1.0 V) & - 2\% & 85.90 (1.2 V) \\
6\% & 365.44  (1.5 V) & - 6\% & 86.77 (0.9 V)\\
10\% & 232.5   (0.2 V) & - 10\% & 81.59   (0.8 V)\\
14\% & 70.32 (1.1 V) & - 14\% & 42.85 (0.1 V)\\
\hline
\end{tabular}
\end{center}
\end{spacing}
\label{tab.2}
\end{table}
In the pristine configuration, the centre-doped APNR clearly transports magnitude of current higher than the other doped ribbons. This is due to the fact that in an H-edge passivated APNR the H-P bonds are stronger and the edge states are located deep in the bands \cite{Wu2015}. The CBM (conduction band minima) and the VBM (valance band maximum) are contributed by the atoms in the central region while centrally doped Si creates dopant state in the Fermi level. Consequently, the major part of the current flows through the central pathways as also observed from the transmission analysis. Henceforth to magnify the current values further, the delocalised states created by the centrally doped Si is preferable compared to edge doped configurations.

\subsection{Influence of Strain}
\begin{figure*}
\includegraphics[scale=0.6]{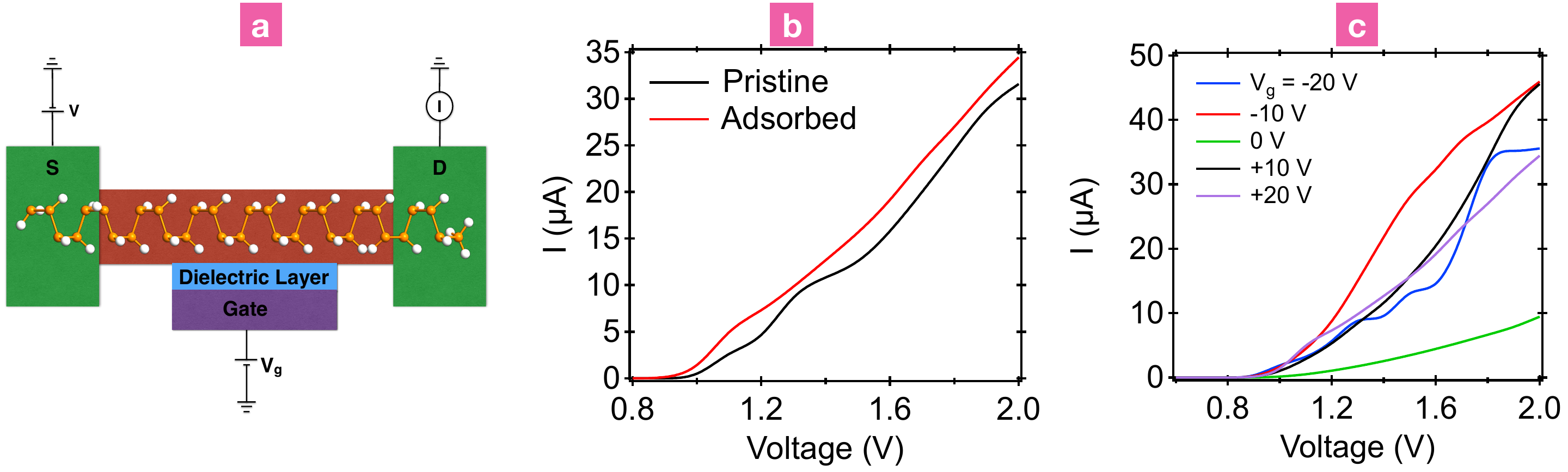}
\caption{{\small (a) APNR based FET structure with a backgate, (b) IV-characteritics of $\mathrm{A_{34}}$ at V$_g$ = 20V in pristine and adsorbed configurations, (c) IV-response in the adsorbed configuration (A$\rm{_{34A}}$) at various values of V$_g$.}}
\label{fig.7}
\end{figure*}
To investigate the influence of external strain on the conductivity of the APNR, electronic and transport properties were studied in the pristine and adsorbed configurations. The percentage uniaxial strain is defined using the following equation : $S = 100\%\times (a - a_0)/a_0$ where $a_0 (a)$ is the lattice constant in the unstrained (strained) configuration of the APNR. In this work, we have covered a significantly large strain range of S = -15\% to +15\%. Within this region of strain, the gap size of the APNR stayed within a range of 0.8 - 1.3 eV (Fig.~S6 in SI\cite{SI-Phosphorene-transport}). In the adsorbed condition, no major changes in the band gap was observed except a small difference at S = 6\%. This suggests that the semiconducting nature of the material do not get affected on benzene adsorption in the strained configuration. The APNR was found to be stable within the elastic limit for a maximum applied strain of S = 30\%.

The semiconducting nature of the APNR is visible from the IV-behaviour of $\mathrm{A_{34P}}$ at $S = 2\%$ as shown in Fig.~\ref{fig.6}(a). On adsorption, the current starts flowing at a much smaller bias compared to $\mathrm{A_{34P}}$ and above 1.45V, the current traces a similar path as that of its pristine counterpart. At $S = 6\%$, the difference in the IV-charactristics in Fig.~\ref{fig.6}(b) between the pristine and adsorbed configurations are much more pronounced over the entire range of applied bias. In the adsorbed configuration, above 0.2V, current starts increasing with an increase in the applied bias with a current peak at 1.36V followed by a minimum at 1.6V - above which it increases sharply at higher voltages. This kind of conductance oscillations have been observed in Phosphorene and other 2D materials \cite{Nguyen2011, Chowdhury2011} which can be explained using the changes in the transmission behaviour as shown in Fig.~S7 in SI\cite{SI-Phosphorene-transport}. Similar peak and valley regions are weakly visible in the pristine configuration, but the overall amplitude of current oscillations is relatively less. In the compressive strain region [Fig.~\ref{fig.6}(c-d)], the gapped electronic structure of the APNR is noticeable from the IV-responses and the current in the adsorbed configuration is lower than the pristine configuration over the entire bias range. This is in contrast to the tensile strain regime where the current gets enhanced on adsorption.

Here, it is observed that with an increase in the applied strain (in tensile and compressive regimes), the APNR starts conducting at a lower voltage than the unstrained APNR in both absorbed and unabsorbed configurations. For S = 6\% non-zero current is measured at 0.2 volt and for both S = 10\% and 14\% conduction started at an astonishingly small bias of 0.1 volt), in spite of having a larger band gap. Similar trend is observed in the compressive strain region, where for S = -6\%, the current starts flowing at 0.1V for the pristine case. This disparity in the relation of large bandgap and low bias voltage for conducting current is the consequence of electrodes becoming semi-metallic when strained. The reason behind this can be explained in the following way. Under the application of an in-plane strain, the layer gets stretched horizontally. This enhances the in-plane bond length between the P-atoms and a reduction of the out-of-plane bond length. As a result, the overlap of 3$p_x, 3p_y$ orbitals between neighbouring P atoms gets reduced, while the 3$p_z$ contribution gets enhanced in respective cases. When the in-plane lattice sites are widely separated, this diminishes the scattering probability by increasing the scattering time due to this reduced orbital overlap. This results in the current hike when the applied strain is increased upto a certain extent. In the compressive strain region, the hopping probability gets enhanced with a higher degree of overlap between the 3$p_x, 3p_y$ orbitals which can explain the onset of the current at lower bias.

The conductivity changes obtained for different strain configuration is listed in Table.~\ref{tab.2} which is significantly large for the entire region of strain studied here. Overall, $\eta$ is  higher in the tensile regime compared to the compressive cases. At S = 2\%, a colossal value of $\eta$ = 1527.8\% was observed which on enhancement of tensile strain stays high with a value of 365.44\% at S = 6\%. Within the Landauer's formalism, such high value of $\eta$ can be explained by comparing the transmission behaviour in the pristine and adsorbed cases as illustrated in Fig. S8 in SI \cite{SI-Phosphorene-transport} for S=2\% at 1V of applied bias. Between the energy range of -0.46 eV$\rightarrow$-1.07 eV and 0.87eV$\rightarrow$ 1.02 eV, transmission is mostly absent in the pristine system while it is non-zero for the adsorbed case. Compared to the pristine case, T(E) is significantly higher for the adsorbed case upto an energy of -1.5 eV. Similar differences in transmission was also observed for S = 6\% at 1.5V of applied bias (Fig. S9 in SI\cite{SI-Phosphorene-transport}) which explains the origin of high value of $\eta$ for these cases. The value of $\eta$ in the compressive strain regime stayed over 80\% upto S = -10\%. Such level of unprecedented conductivity modulation has not been observed in Phosphorene earlier which suggests that the application of strain can be a very effective way of achieving enhanced conduction in Phosphorene.

\subsection{Effect of a Gate Voltage}

To understand the influence of gate voltage on the conductance behaviour, the transport properties of APNR was studied in a FET geometry as illustrated in Fig.~\ref{fig.7}(a). The structure consists of a metallic gate connecting the APNR through a dielectric layer. The dielectric constant was taken as $\epsilon_r$ = 4.2 which is similar to that of SiO$_2$, commonly used in FETs. In this all Phosphorene FET, the Source (S) and Drain (D) electrodes are also considered to be made of Phosphorene which rules out the presence of barriers at the metal-semiconductor interfacial contacts and their effects on transport properties. 

In the presence of a $V_g$, the IV-pattern retains the semiconducting character of the APNR in the FET structure as can be seen in Fig.~\ref{fig.7}(b). At V$_g$ = 20V, pronounced difference in the current can be observed between the pristine and the adsorbed configuration that resulted in an extreme $\eta\sim$300\% which at V$_g$ = 10V also stays very high at 200\%. Within a large V$_g$ range of -20V to +20V, the IV-characteristics in the adsorbed configurations are illustrated in Fig.~\ref{fig.7}(c). With an increase in the bias, the difference in the current gets enhanced which is maximum at V$_g$ = 10V, -10V with a maximum $\rm{\Delta I\sim 36.5}\mu$A. At higher values of V$_g$ = 20V, -20V, it is also significantly large at 26$\mu$A at 2V bias. Firstly, the presence of such large current ($\sim 45 \mu$A) at both positive and negative values of V$_g$ confirms that the ambipolar nature of the APNR stays intact in the adsorbed configuration. Secondly, the transverse electric field of the gate voltage offers a better enhancement of current in the adsorbed configuration compared to its pristine counterpart. This indicates the effect of the presence of Benzene molecule on the conduction behaviour of APNR. The large change in the current due to a change of V$_g$ suggests that a backgate can  control the conduction response in a reversible manner without bringing permanent changes in the material.

Apart from the configuration discussed in this work, the conductance modulation was observed irrespective of the adsorption configuration considered from top or side of the nanoribbon. This effect is present under the influence of various external influences like (a) application of strain, (b) presence of dopants of various types at different locations and (c) inclusion of a gate voltage, supporting not only the robustness of this conductivity enhancement, but as well further tuneability obtained through the variation of such parameters suggest the global nature of this claim. Based on the trend observed, it is expected that multiple adsorption of molecules will lead to a much larger change in conductivity in such APNR.

As the width of APNR grows, the changes in conductance could be reduced, however, the effect is expected to smear only for large flakes with micron width range, while for nanoribbons of technological interest having few nm width, it still expects the changes to be significant. Experimentally, nanoribbons of similar dimension in 2D crystals have been grown already using multiple routes \cite{Li2009}, in lateral heterostructure \cite{Sutter2012}, vertical stacks \cite{Liu2011} and from a liquid precursor \cite{Srivastava2010}. Introduction of nanoscale defects \cite{Moreno2018, Ziatdinov2013}with atomic scale uniformity and long-range order in such 2D structures have been demonstrated. In addition, STM-based lithography techniques \cite{Tapas2008, Sutter2012, Cyr1996} have been useful in generating preferred edge structure and detection of attached molecules in a nanoribbon. Strain can be applied using a bending apparatus, nanoindentation in an atomic force microscope or through the use of flexible substrates \cite{Ray2018}. Single molecular detection has been experimentally observed for graphene \cite{Schedin2007, Mohanty2008}, Carbon Nanotube \cite{Guo2006, Goldsmith2007} where decreasing the channel length led to an increase in the sensitivity upto a critical dimension as considered in the present work. These reports suggest possibilities to create nanoribbons of current interest, especially with STM based local oxidation. Such prospect makes our work as a useful guideline for future experiments.

\section{Conclusion}
In this work, First-principles based NEGF-DFT calculations are used to investigate the electronic and transport properties of a Phosphorene nanoribbon in the close proximity of an organic molecule and detected a supreme modulation of the conduction response. We observed that in the absence of any external stimuli, the conduction behaviour changes significantly in the presence of the molecule while the semiconducting nature of Phosphorene stays similar. The conduction modulation has been characterised by a modulation parameter ($\eta$) which has a maximum $\sim$ 150\% in this environment. We observed that the behaviour can be further controlled by (i) doping, (ii) applying strain and (iii) gate voltage. Independent of the doping type, location and concentration - a paramount amount of change was observed with a maximum $\eta\sim$ 570\% with edge doped Sulphur. On the other hand under the application of strain, a colossal enhancement of current and $\eta\sim$ 1528\% was measured under 2\% tensile strain. The gate voltage offers a controlled tuning of this effect with a maximum conduction change $\eta\sim$ 300\% in non-invasive manner. Due to the larger values of measured currents (20-50$\mu$A) and a gate controlled  conductivity modulation of almost 400\% ($\mathrm{\Delta V_g = 10V}$) in the adsorbed configuration, the feasibility of practical studies are significantly high thanks to the advancement of the present day nanotechnological tools. The conductivity modulation will be prevalent for nanoribbons with few nm size as commonly required for the fabrication of a semiconducting channel in a present-day electronic device, compared to that of a much wider nanoribbon with larger numbers of  central transport pathways.Our current findings provide new insight about proximity induced tuning and control of the conduction response of Phosphorene and propose new direction for its usefulness in various nanoscale applications in switching, superior sensing\cite{Ray2015a, Ray2015b} and energy efficient designs etc.

\section*{Acknowledgments}

This work was financially supported by Department of Science and Technology, Govt. of India through the INSPIRE scheme (Grant Ref: DST/INSPIRE/04/2015/003087).

% %  Bibliography

\bibliographystyle{unsrt}

\end{document}